\documentclass[aps,twocolumn,eqsecnum]{revtex4}

\usepackage{natbib}
\usepackage{graphicx}
\usepackage{amsmath}
\usepackage{amssymb}
\usepackage{subfig}

\newcommand\oneunit[1]{\nobreak\mbox{$\;$#1}}
\newcommand\permcub{\nobreak\mbox{$\;$m$^{-3}$}}
\newcommand\tento[1]{\nobreak\cdot\nobreak10^{#1}}

\def\newblock{\hskip .11em \@plus .33em \@minus .07em}

\begin{document}

\title{Particle acceleration in a reconnecting current sheet: PIC simulation}



\author{T.\,V. Siversky and V.\,V. Zharkova}

\affiliation{Computing Department, University of Bradford, Bradford BD7 1DP, UK}




\begin{abstract}
The acceleration of protons and electrons in a reconnecting current sheet (RCS) is simulated with a particle-in-cell (PIC) 2D3V code for the proton-to-electron mass ratio of 100. The electro-magnetic configuration forming the RCS incorporates all three components of the magnetic field (including the guiding field) and a drifted electric field. PIC simulations reveal that there is a polarisation electric field that appears during acceleration owing to a separation of electrons from protons towards the midplane of the RCS. If the plasma density is low, the polarisation field is weak and the particle trajectories in the PIC simulations are similar to those in the test particle (TP) approach. For the higher plasma density the polarisation field is stronger and it affects the trajectories of protons by increasing their orbits during acceleration. This field also leads to a less asymmetrical abundances of ejected protons towards the midplane in comparison with the TP approach. For a given magnetic topology electrons in PIC simulations are ejected to the same semispace as protons, contrary to the TP results. This happens because the polarisation field extends far beyond the thickness of a current sheet. This field decelerates the electrons, which are initially ejected into the semispace opposite to the protons, returns them back to the RCS, and, eventually, leads to the electron ejection into the same semispace as protons. Energy distribution of the ejected electrons is rather wide and single-peak, contrary to the two-peak narrow-energy distribution obtained in the TP approach. In the case of a strong guiding field, the mean energy of the ejected electrons is found to be smaller than it is predicted analytically and by the TP simulations. The beam of accelerated electrons is also found to generate turbulent electric field in a form of Langmuir waves.
\end{abstract}

\maketitle

\section{Introduction} \label{sec:intro}

The observations in hard X-rays and gamma-rays of solar flares or geomagnetic tail imply that the essential part of the released energy has to be converted into accelerated particles -- electrons and ions. It is usually assumed that the acceleration occurs as a result of conversion of the free magnetic energy released in a magnetic reconnection. The following three mechanisms of particle acceleration are normally considered: acceleration by an electric DC field \citep{Litvinenko93}, stochastic acceleration of different kinds \citep{Miller96} and shock acceleration \citep{Cargill88}. However, it is not still clear which of the mechanisms of particle acceleration really takes place in solar flares or geomagnetic tail \citep{Miller97, Aschwanden02}. Each mechanism can be found contributing to the emitted radiation and in order to distinguish their effects more accurate simulations of particle trajectories and energy spectra of accelerated particles are required.

In the present study we focus on particle acceleration by electric field inside a reconnecting current sheet. This field can be considered as a logical step in progressing with a conversion of the magnetic field energy into the energy of accelerated particles. The first analytical study of this acceleration mechanism was proposed by \citet{Speiser65} who found that the particles can be accelerated to rather high energies after they enter the configuration with reconnecting magnetic field lines forming an RCS and a perpendicular (drifted) electric field. This theory was extended by \citet{Litvinenko93, Litvinenko96} who considered an RCS with all three components of the magnetic field and carried out analytical estimations for the energy of accelerated particles. In particular, the authors found that the guiding magnetic field can significantly increase the energy of accelerated particles.

\citet{Efthym05} have analytically studied the trajectories of particles inside a similar current sheet configuration by means of the dynamical systems methods. In particular, they found that in a certain magnetic configuration a particle can be trapped inside the RCS. Conditions of a particle trapping have been found in case of a weak guiding magnetic field, while for a strong guiding field particles always follow escaping orbits. The energy gain for escaping particles is determined as a function of initial position and velocity.

Particle trajectories in the similar 3D magnetic configuration with the guiding field were further studied numerically by using a test particle approach by \citet{Zharkova04} (constant electric field) and \citet{Wood05} (electric field enhanced near the X-nullpoint due to anomalous resistivity). \citet{Zharkova04} showed that the trajectories of particles with the opposite charges (electrons or protons) can be either fully symmetric or strongly asymmetric towards the midplane of the RCS depending on the ratio between the magnetic field components. If the guiding field is strong enough, the accelerated electrons and ions are found ejected into the opposite directions with respect to the midplane causing charge separation, if the guiding field is weak, the particle are ejected in even proportions into each side from the midplane. The direction of ejection is shown to be dependent on the directions of the transverse and guiding magnetic fields as well as the direction of the drifted electric field. It was also shown that the accelerated particles gain energies up to 100\oneunit{keV} for the electrons and up to 1\oneunit{MeV} for the protons for the electric field magnitude of 100\oneunit{V/m} \citep{Zharkova05mn, Wood05}.

As a result, the energy spectra of accelerated particles also depend on a magnetic field topology, an electric field strength and the dependence of transverse magnetic field variations on a distance from the X-nullpoint. For example, the spectral indices of energy spectra vary from 2 for electrons and 1.5 for protons if the transverse magnetic field linearly increases with the distance, or they change for a particular magnetic model adopted from the MHD simulation of \citet{Somov00} to 1.8-2.2 for electrons and 1.3-1.8 for the protons. In the similar model \citet{Wood05} obtained for electrons a power-law spectra with index $\approx 1.5$. The all spectral indices for both electrons and protons become much higher if the transverse magnetic field increases exponentially with a distance from the X-nullpoint \citep{Zharkova05}.

Therefore, particle trajectories gained during acceleration in an RCS is usually considered non-self-consistently, i.e. in modelled electric and magnetic fields. Such studies can give some information about accelerated particles but they do not include fields generated by the accelerated particles. In the previously mentioned papers the current sheet was assumed to be already formed and remained stationary while reconnecting.  On the other hand, there are also studies where magnetic reconnection is treated in a framework of kinetic (self-consistent) simulation \citep[see, e.g.][]{Shay98, Birn01, Drake06, Tsiklauri07}. Most of this studies are focused on the reconnection itself: determining the rate of reconnection and the outflow velocity, studying the structure of the dissipation region.

However, only in recent studies the PIC simulations were applied to consider particle acceleration during a magnetic reconnection in the vicinity of X-nullpoint. \citet{Tsiklauri07} used PIC simulation to study the whole evolution of reconnecting magnetic field which started from the stressed X-nullpoint configuration and evolved to the unstressed one within a few thousands of Alfv\'en times. While the main goal was to investigate the reconnection processes in the vicinity of X-nullpoint in a compressed current sheet, the authors show that some fraction of the the released magnetic energy is transformed into the kinetic energy of accelerated particles. The energy spectrum of accelerated electrons is power-law with the index in a range from 4 to 5.5 and maximal energy from 100\oneunit{keV} to 2\oneunit{MeV} depending on the stress factor of the X-nullpoint. Another model of particle acceleration was also studied by using the PIC simulation by \citet{Drake06} in the reconnecting magnetic configuration called magnetic islands. The PIC simulation reveals that the multiple interactions of electrons with these magnetic islands allow them to reach relativistic energies. However, there is no evidence that such mechanism can be effective for ions.

In order to study particle acceleration in a current sheet in the current paper we will also use the PIC approach, which takes into account electric and magnetic fields generated by accelerated particles. Similar to the other authors, we do not simulate the whole reconnection site  but a relatively small part of the RCS at some distance from the X-nullpoint with the size large enough to contain a whole particle orbit before its ejection. Similar to \citet{Drake06}, we study the 3D current sheet which is already formed by a reconnection process. The RCS configuration in our study is similar to one discussed by \citet{Litvinenko93, Zharkova04}. Since the time scale of particle acceleration is much smaller than the typical MHD time scales, we assume that the background magnetic field which forms the RCS does not evolve during the particle acceleration.

Although it has to be noted that PIC simulations for space plasmas have a few essential problems \citep[see for the full details introduction in][]{Zharkova08}. Probably, the main problem for the explicit PIC simulations is a limited size of the simulation region restricted by a modern computer power. This problem is the result of a condition that the step of the spatial grid cannot exceed the Debye length $\lambda_D = (k T / 4 \pi n e^2)^{1/2}$. For example, in the solar corona the Debye length is of the order of $10^{-3}\oneunit{m}$. Thus, the simulation regions in the PIC simulations of the magnetic reconnection carried out by \citet{Birn01, Tsiklauri07} in the solar corona conditions would have sizes of the order of few metres at the most. Therefore, it is very doubtful that PIC simulations can be used to study large scale reconnection events often occurring in the Sun \citep{Priest81}.

Much more realistic approach is to use the PIC simulation to study small scale objects, with the sizes of the order of a current sheet thickness. For example such the study was carried out by \citet{Drake06} for the PIC simulation of two magnetic islands within a current sheet. However, the size problem discussed above leads to the following restrictions. For the current sheet thickness equal to the ion inertial length, $\delta_i$, the number of cells across the current sheet in a PIC simulation have to be $\delta_i/\lambda_D = c \sqrt{m_i/(kT)}$, which is $3\tento{3}$ for the solar corona temperature or $3\tento{4}$ for the magnetosphere. In order to reduce this number, \citet{Drake06} used a reduced magnitude for the speed of light, $c=20V_A \approx 6\tento{6} \oneunit{m/s}$, where $V_A$ is the Alfv\'en velocity in the magnetosphere. Another way to reduce the number of cells was used in the PIC simulation carried out by \citet{Karlicky08}, who considered the high temperature electron-positron plasma, for which the ratio $\delta_i/\lambda_D$ was as low as $10$.

We assume that the current sheet is formed by the plasma with the typical coronal density of $10^{16}\permcub$, which sustain the equilibrium with the background magnetic field. In order to avoid the problem with the small Debye length, only a small fraction of the plasma particles (with density of $10^{10}\permcub$) is included in the PIC simulation. This makes the ratio $\delta_i/\lambda_D$ to be of the order of $10$. This approach allows us to have a reasonable number of cells and super-particles without using the reduced speed of light or ions with very small mass. Also, this approach can be considered as a development of the TP method towards the full particle simulation. As it is shown further, even such a small number of simulated particles can demonstrate the new effects, which are not possible in the TP approach, for example, they can generate a strong polarisation electric field. We also use the TP approach to determine how the locally induced fields affect the particle trajectories.

\section{The model} \label{sec:model}

\begin{figure}
  \centering
  \includegraphics[width=0.49\textwidth]{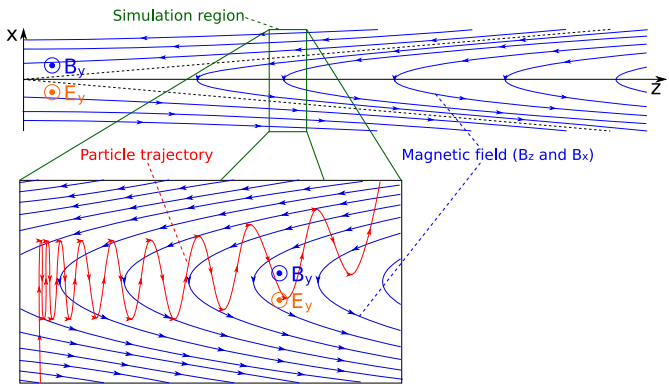}
  \caption{Simulation model.}
  \label{fig:sim_mod}
\end{figure}

The model accepted in this paper (Fig.~\ref{fig:sim_mod}) is a further development of the test particle model used by \citet{Zharkova04}. We consider the reconnecting magnetic field as a background field, which is caused by the external processes of magnetic reconnection and intend to simulate with the PIC approach the particle trajectories and their density/energy spectra in this field. Contrary to the TP simulations, plasma particles in the PIC simulations are considered to generate their own electric and magnetic fields, which is now self-consistently taken into account.

TP simulations have shown that the acceleration time is of the order of $10^{-5}\oneunit{s}$ for the electrons and $10^{-3}\oneunit{s}$ for the protons. Since this time is much shorter than the time of the reconnecting magnetic field variation \citep{Priest00}, then we can assume that during simulation the background magnetic field is stationary. Also, from the TP simulations we conclude that travel distances of accelerating particles along the RCS are of the order of $10\oneunit{km}$ at most (for the protons) \citep{Zharkova08}. Thus we can assume that this length is much shorter than the length scale of the magnetic field variation along the current sheet. In addition, as it is generally accepted, we suppose that the magnetic field variations across the current sheet has much shorter length scale than its variation along the current sheet,
\begin{equation}
    L_x \ll L_z, L_y.
\end{equation}

Thus, our simulation domain is a small part of the RCS (see Fig.~\ref{fig:sim_mod}), large enough to contain the full trajectories of accelerated particles. The background magnetic field is stationary and vary inside this domain only in the $x$ direction, which is perpendicular to the RCS. We take into account all three components of the background magnetic field. The main component $B_z$ depends on $x$ as follows:
\begin{equation}
    \label{eq:bz}
    B_z(x) = -B_{z0} \tanh\left( \frac{x}{L_x} \right).
\end{equation}
The $B_x$ component is assumed \citep[like in][]{Zharkova04} to be constant inside the simulation domain:
\begin{equation}
    \label{eq:bx}
    B_x = -B_{x0}.
\end{equation}
The guiding (out-of-plane) magnetic field $B_y$ is maximal in the midplane and vanishes outside the RCS:
\begin{equation}
    \label{eq:by}
    B_y(x) = B_{y0} \mathrm{sech}\left( \frac{x}{L_x} \right).
\end{equation}
Note, that if $B_{y0} = 0$ the configuration corresponds to the Harris sheet equilibrium, and if $B_{y0} = B_{z0}$ the equilibrium is force-free.

The inflow of plasma into the RCS combined with the condition of the frozen-in magnetic field leads to the induction of the drifted (out-of-plane) electric field $E_y$. In order to provide the inflow of plasma in our simulation domain we set up a background electric field, as those drifted in with velocity $V_{in}$ by a magnetic diffusion process \citep{Priest00}.
\begin{equation}
    \label{eq:ey}
    E_y = E_{y0} = B_{z0} V_{in},
\end{equation}
where $V_{in}$ is the inflow velocity.

In our study we use the following values for the current sheet parameters: the main component of the magnetic field $B_{z0}=10^{-3} \oneunit{T}$, the current sheet half-thickness $L_x = 1 \oneunit{m}$, the drifted electric field $E_{y0} = 250 \oneunit{V/m}$ and the guiding, $B_{y0}$, and transverse, $B_{x0}$, components of the magnetic field are varied to study how they influence particle acceleration.

\section{Test particle simulations}

\begin{figure}
  \centering
  \includegraphics[width=0.48\textwidth]{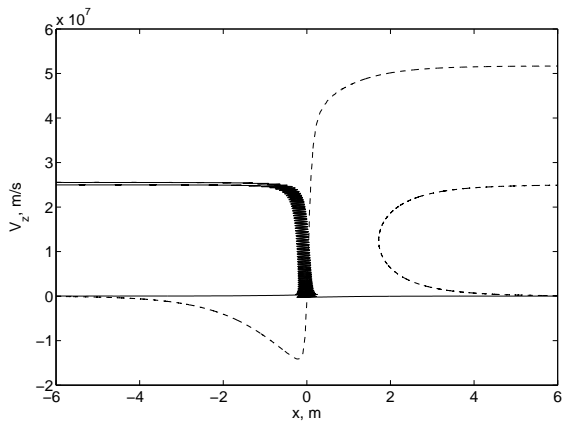}
  \caption{Trajectories of two protons (solid) and two electrons (dashed) on the $x$-$V_z$ phase plane during acceleration inside a current sheet (TP simulations). $B_{y0} = 10^{-4} \oneunit{T}$, $B_{x0} = 2 \tento{-5} \oneunit{T}$, $m_p/m_e=100$.}
  \label{fig:vz_x}
\end{figure}

For better understanding of particle trajectories inside an RCS during the PIC simulations let us first investigate the results obtained in the TP approach \citep[similar to those by][]{Zharkova04}.

A typical trajectory of a plasma particle (Fig.~\ref{fig:vz_x}) in electro-magnetic configuration shown in Fig.~\ref{fig:sim_mod} consists of the three parts: $\mathbf{E} \times \mathbf{B}$ drift, acceleration by the electric field and ejection. (i)Outside the current sheet, where the magnetic field is strong, particle motion is adiabatic and can be described as the superposition of a magnetic gyration and a drift in the orthogonal electric and magnetic fields towards the midplane of the RCS. (ii)Inside the current sheet the particle moves along the $y$ axis and is accelerated by the electric field $E_y$. The exact trajectory of the particle depends on the value of guiding field $B_y$ \citep{Litvinenko93}. For a large $B_y$ the particle remains magnetised and reaches higher energy than for a small $B_y$, when the particle becomes unmagnetised inside the current sheet. (iii)When the particle gains enough energy the magnetic configuration cannot confine the particle motion any more and it is ejected from the RCS. Outside the current sheet the particle motion is again adiabatic. However, due to a high velocity of the particle along the magnetic field line and non-zero $B_x$ the $x$ component of the particle velocity is larger than the $\mathbf{E} \times \mathbf{B}$ drift velocity and the particle moves away from the RCS.

Fig.~\ref{fig:vz_x} shows the trajectories in the $x$-$V_z$ phase plane of two protons and two electrons (one proton and one electron inflow from the $x>0$ semispace another pair from the $x<0$ semispace). The mass ratio is chosen to be $m_p/m_e=100$ for the comparison with the PIC simulations in further sections. The magnitudes of the magnetic field components $B_y$ and $B_x$ are chosen such that electrons remain magnetised in the vicinity of the midplane while protons are unmagnetised during the acceleration phase. In this case the energy of the unmagnetised protons upon ejection can be estimated \citep[see][]{Litvinenko96} as
\begin{equation}
    \label{eq:enrg_x_unmag}
    \varepsilon = 2 m_{\alpha} \left(\frac{E_{y0}}{B_{x0}}\right)^2,
\end{equation}
where $m_{\alpha}$ is the particle mass. As it was shown by \citet{Zharkova04} if $B_{y0}$ is strong enough ($>1.5 \tento{-2} B_{z0}$) all protons (regardless of side they entered from) are ejected to one semispace while all electrons are ejected to the opposite semispace with respect to the midplane. Indeed, in Fig.~\ref{fig:vz_x} both protons are ejected to the semispace with negative $x$ and both electrons to the $x>0$ semispace. In order to distinguish two types of trajectories, the particles that inflow from and are ejected into the same semispace will be referred to as "bounced" particles (this corresponds to the proton coming from $x<0$ and electron coming from $x>0$ in Fig.~\ref{fig:vz_x}). Particles that are ejected into the opposite semispace from the one they come from will be referred to as "transit" particles (this corresponds to the proton coming from $x>0$ and electron coming from $x<0$ in Fig.~\ref{fig:vz_x}).

As it is seen in Fig.~\ref{fig:vz_x}, "transit" and "bounced" particles gain different energy, and this effect is stronger for electrons. Note, that the difference in motion of "transit" and "bounced" particles vanishes when $B_y \rightarrow 0$ \citep[see also Figs. 6,7 in][]{Zharkova08}. On the other hand, if $B_x = 0$ particles cannot escape from the RCS and their acceleration is only limited by the size of the current sheet in the $y$ direction.

Since for the magnitude of $B_y$ and $B_x$ which were used in Fig.~\ref{fig:vz_x} the electrons are magnetised, then as it was shown by \citet{Litvinenko96} they follow magnetic field lines and gain the energy:
\begin{equation}
    \label{eq:enrg_x_mag}
    \varepsilon = 2 L_x \left| e E_{y0} \frac{B_{y0}}{B_{x0}} \right|.
\end{equation}
However, this relation is only valid for the "transit" magnetised electron. The "bounced" electron, on the other hand, experiences a repelling force caused by $E_y$ while travelling along the magnetic field line towards the midplane. When this force overcomes the attraction due to the $\mathbf{E} \times \mathbf{B}$ drift the "bounced" electron is getting ejected. Since it cannot reach the midplane the "bounced" electron gains less energy than the "transit" one.

\begin{figure}
  \centering
  \subfloat[$B_{y0} = 10^{-4} \oneunit{T}$]{
    \includegraphics[width=0.48\textwidth]{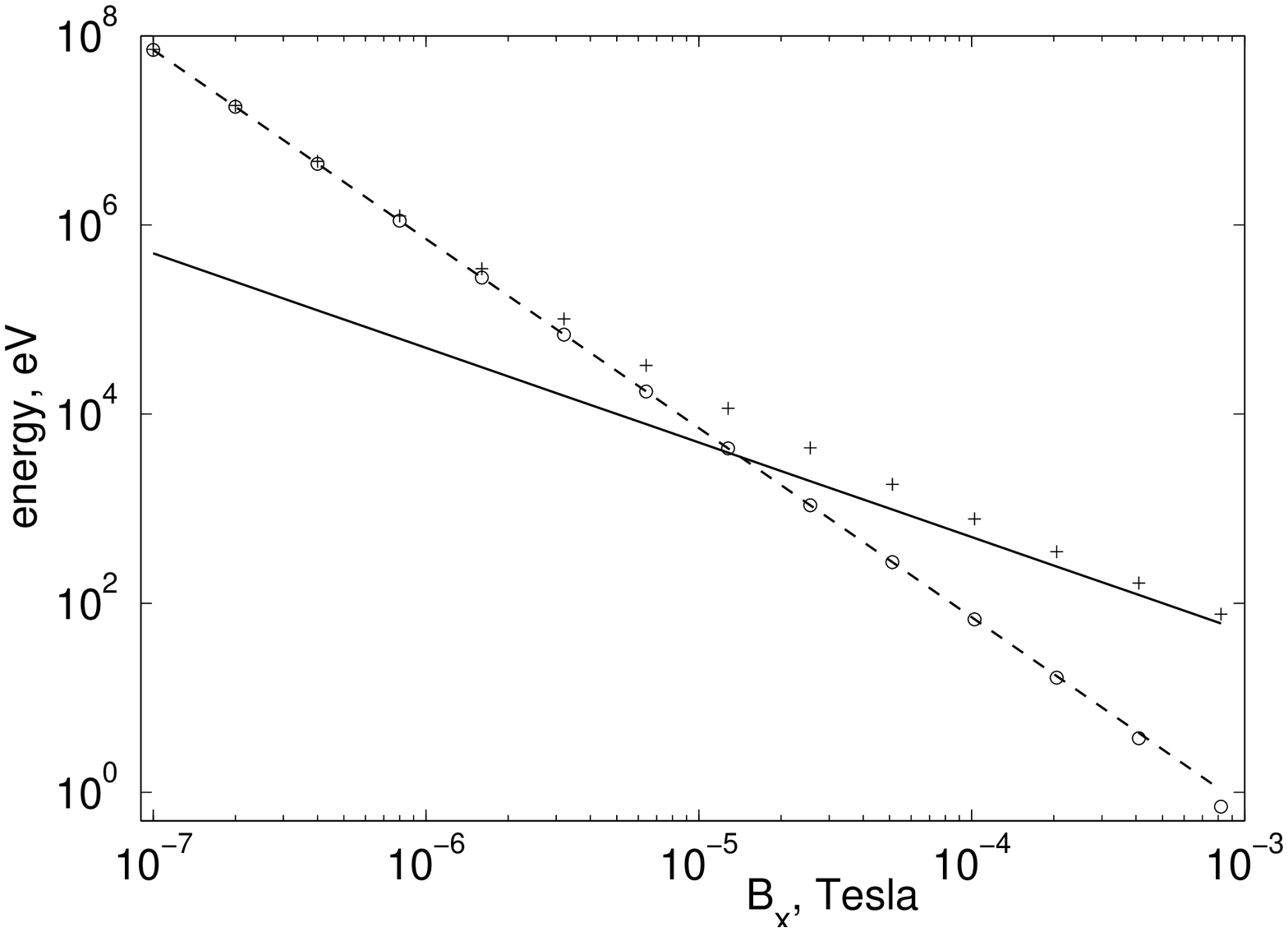}
    \label{fig:enrg_bx}}\qquad
  \subfloat[$B_{x0} = 2 \tento{-4} \oneunit{T}$]{
    \includegraphics[width=0.48\textwidth]{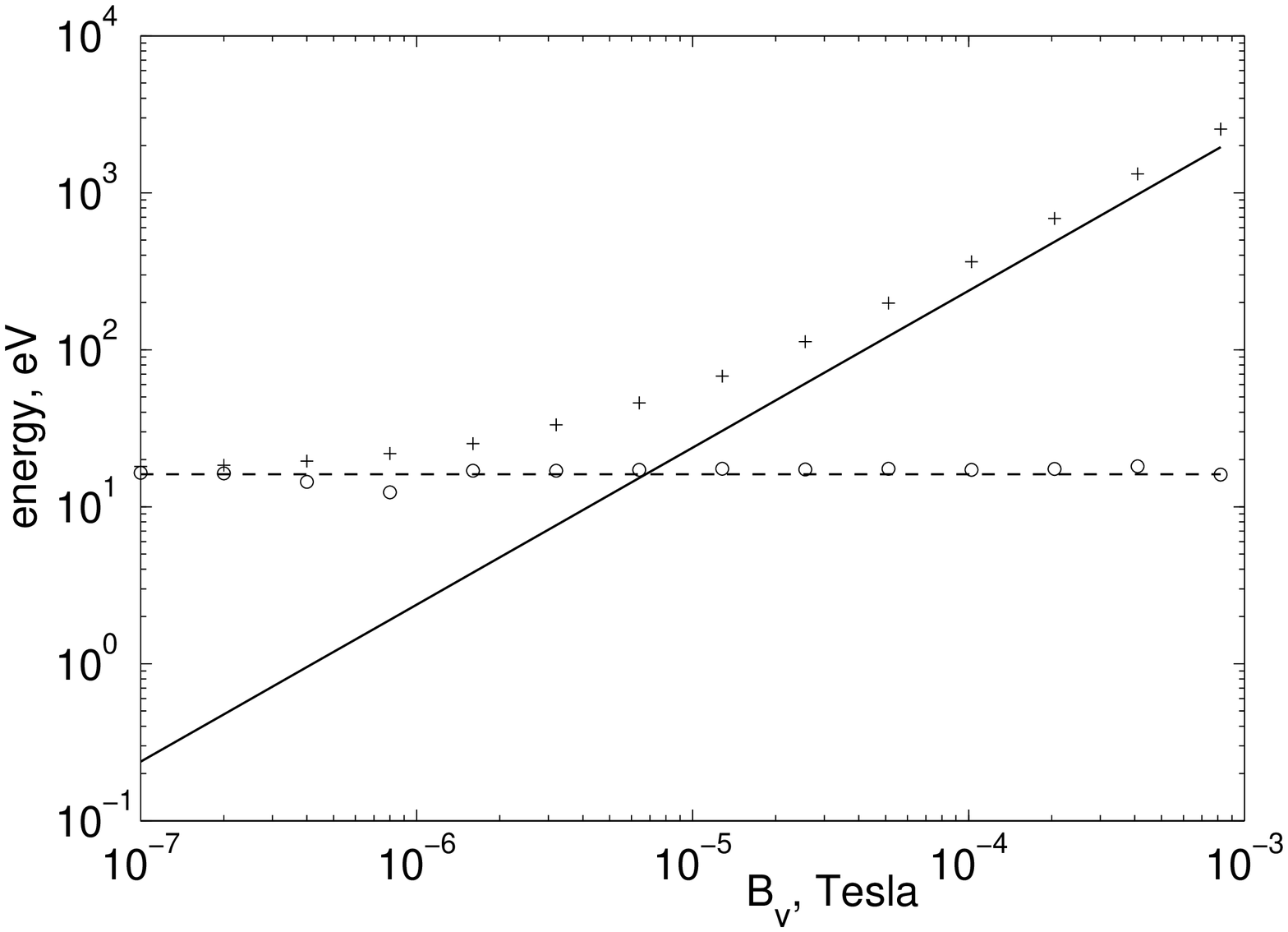}
    \label{fig:enrg_by}}
  \caption{Electron energy in the TP simulations upon ejection as a function of (a)$B_{x0}$ and (b)$B_{y0}$. Crosses -- "transit" electron, circles -- "bounced" electron, dashed line -- Eq.~(\ref{eq:enrg_x_unmag}), solid line -- Eq.~(\ref{eq:enrg_x_mag}). $B_{z0} = 10^{-3} \oneunit{T}$, $E_{y0} = 250 \oneunit{V/m}$, $L_x = 1 \oneunit{m}$.}
\end{figure}

The ejection energies of electrons are plotted in Fig.~\ref{fig:enrg_bx} and \ref{fig:enrg_by} as a function of $B_{x0}$ and $B_{y0}$, respectively. The lines correspond to Eq.~(\ref{eq:enrg_x_unmag}) (dashed) and Eq.~(\ref{eq:enrg_x_mag}) (solid) obtained analytically by \citet{Litvinenko96} in the limits of weak and strong $B_y$. The energies of "transit" electrons are in a good agreement with the analytical estimations, i.e. with Eq.~(\ref{eq:enrg_x_unmag}) for weak $B_y$ and $B_x$ when electrons are unmagnetised, and with Eq.~(\ref{eq:enrg_x_mag}) for strong $B_y$ and $B_x$ when electrons are magnetised. On the other hand, energies of the "bounced" electrons for any $B_y$ and $B_x$ coincide with these of the unmagnetised ones given by Eq.~(\ref{eq:enrg_x_unmag}). This means that $B_y$ enhances the ejection energy of the "transit" electrons only.

\section{Particle-in-cell simulations}

\subsection{Problem formulation}

\begin{figure*}
  \centering
  \subfloat[Protons, $n=10^{7} \permcub$]{
    \includegraphics[width=0.48\textwidth]{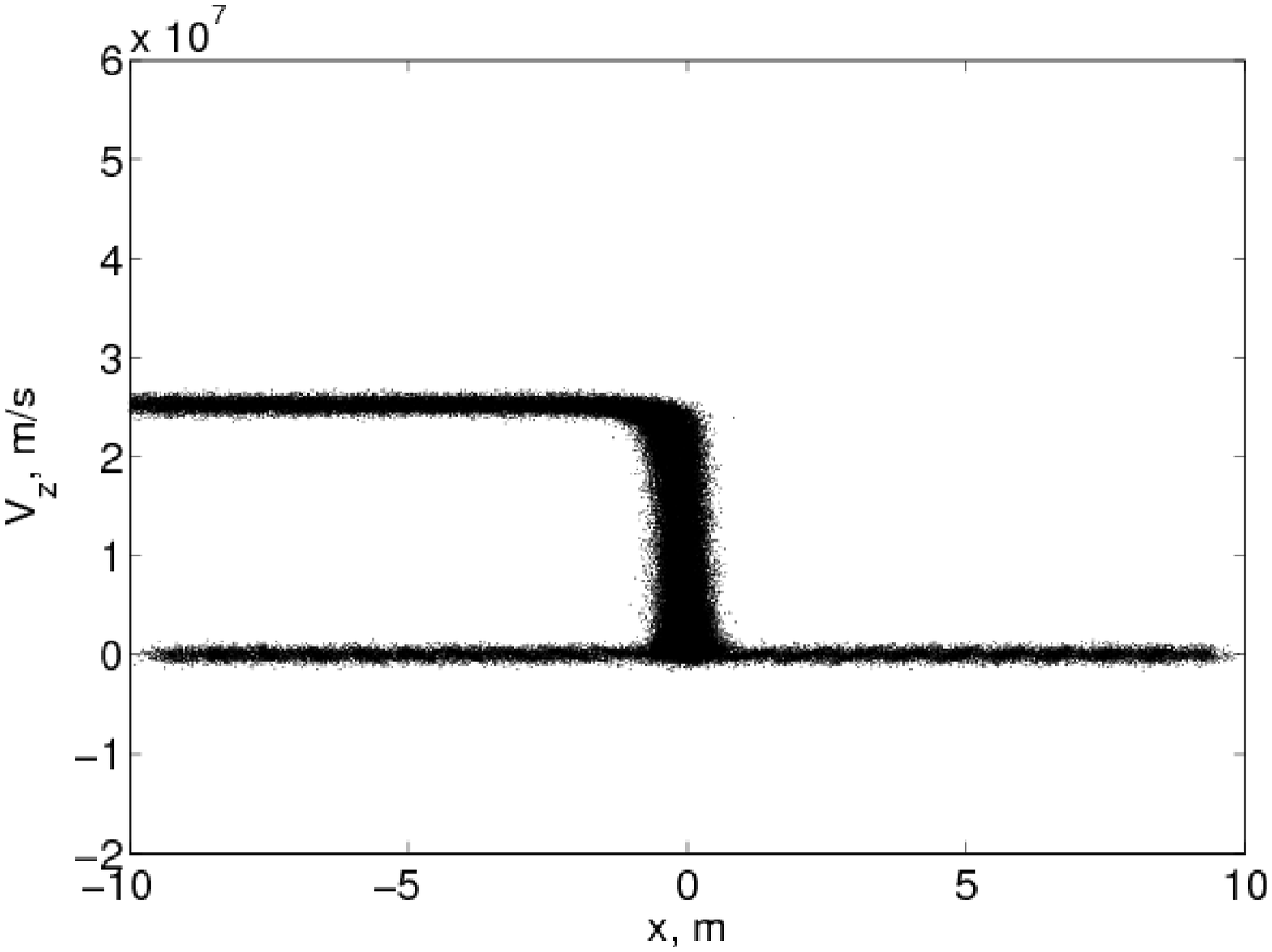}
    \label{fig:vz_x_lownp}}
  \subfloat[Electrons, $n=10^{7} \permcub$]{
    \includegraphics[width=0.48\textwidth]{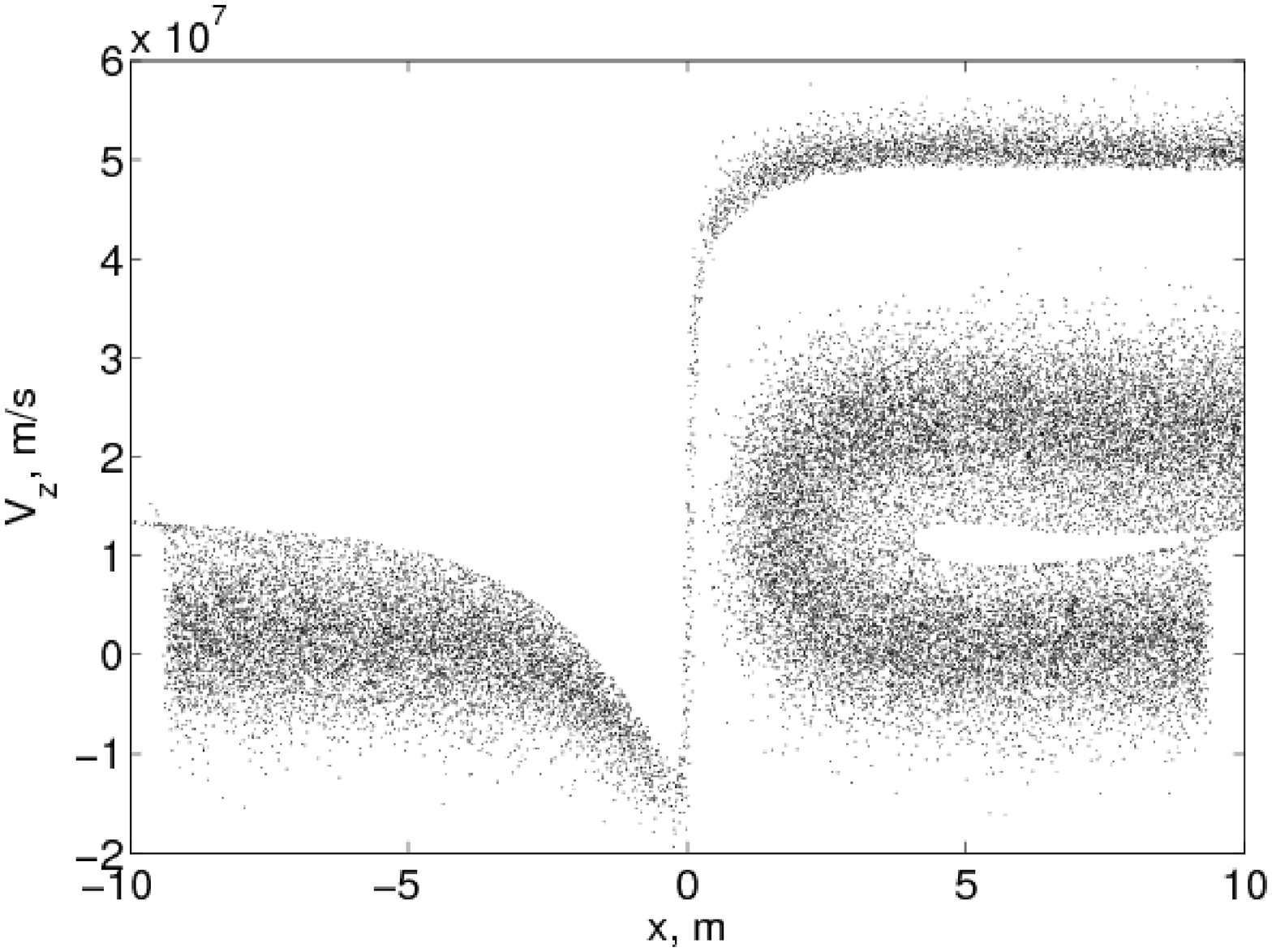}
    \label{fig:vz_x_lowne}}\\
  \subfloat[Protons, $n=10^{10} \permcub$]{
    \includegraphics[width=0.48\textwidth]{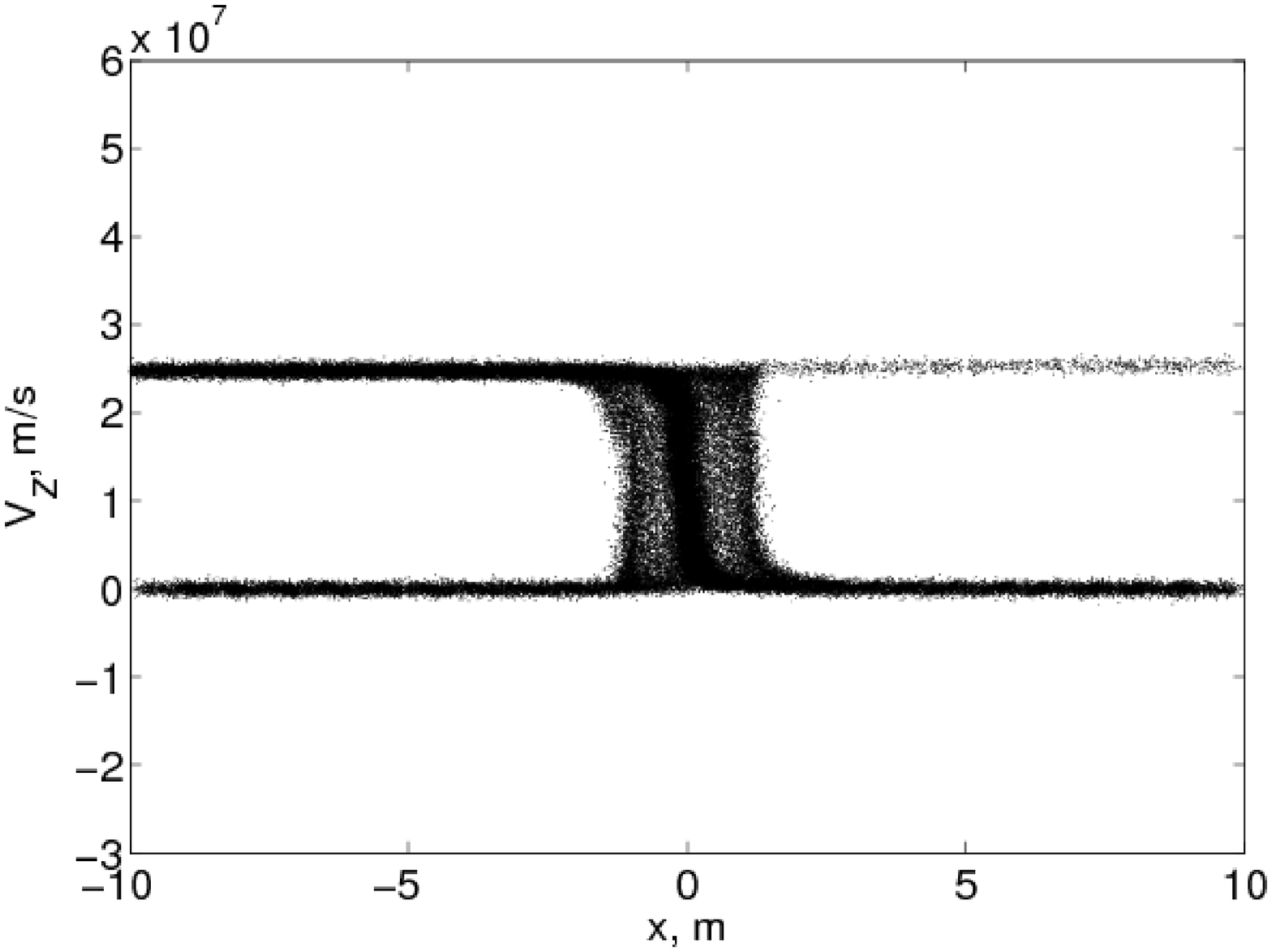}
    \label{fig:vz_x_hinp}}
  \subfloat[Electrons, $n=10^{10} \permcub$]{
    \includegraphics[width=0.48\textwidth]{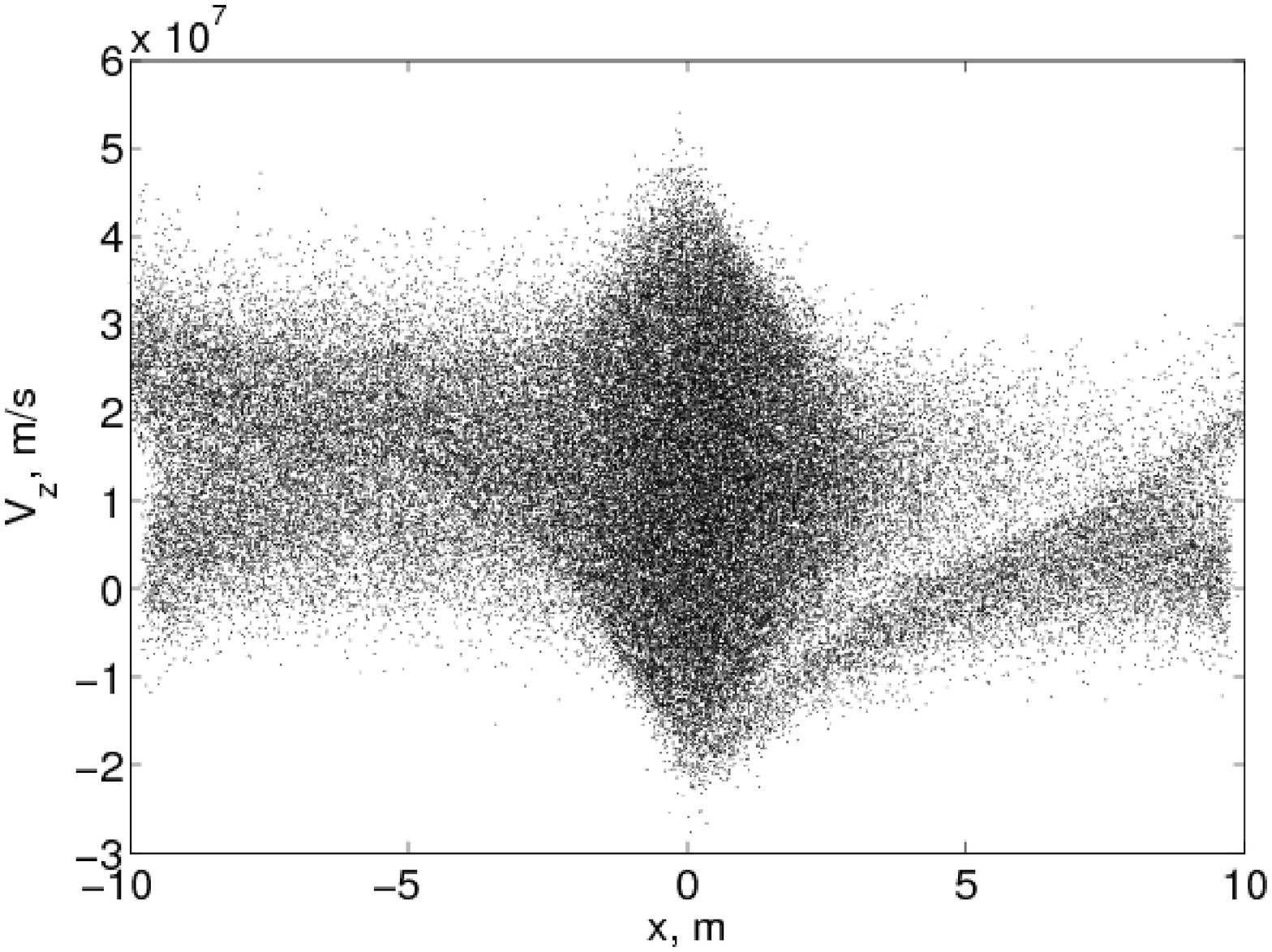}
    \label{fig:vz_x_hine}}
  \caption{Snapshots of the plasma particles on the $x$-$V_z$ phase plane (PIC simulations). Current sheet parameters are the same as in Fig.~\ref{fig:vz_x}.}
\end{figure*}

Although the TP simulations can provide some valuable results of particle motion inside an RCS, it does not take into account the electric and magnetic fields generated by accelerated particles in the simulation region. In order to include these fields, we used 2D3V particle-in-cell simulation code developed by \citet{Verboncoeur95}. PIC method, similar to TP, is based on the equation of motion for plasma particles
\begin{equation}
    \label{eq:motion}
    \frac{d \mathbf{V}_{si}}{dt} = \frac{q_s}{m_s} \left( \mathbf{E}+\mathbf{\tilde{E}}+\mathbf{V}_{si}\times\left(\mathbf{B}+\mathbf{\tilde{B}}\right) \right),
\end{equation}
where besides of the background fields $\mathbf{E}$ and $\mathbf{B}$ given by Eqs.~(\ref{eq:bz})-(\ref{eq:ey}), the local self-consistent fields $\mathbf{\tilde{E}}$ and $\mathbf{\tilde{B}}$ induced by the accelerated particles are taken into account. These fields are calculated from the Maxwell equations:
\begin{eqnarray}
    \label{eq:maxwell1}
    \frac{\partial \mathbf{\tilde{E}}}{\partial t} &=& c^2 \nabla \times \mathbf{\tilde{B}} - \frac{1}{\epsilon_0} \left( \mathbf{\tilde{j_e}}+ \mathbf{\tilde{j_p}} \right), \\
    \label{eq:maxwell2}
    \frac{\partial \mathbf{\tilde{B}}}{\partial t} &=& -\nabla \times \mathbf{\tilde{E}},
\end{eqnarray}
where $\mathbf{\tilde{j_e}}$ and $\mathbf{\tilde{j_p}}$ are the current densities of the electrons and protons.

In our 2D simulations the $y$ dimension is chosen to be invariant. The system is periodic in the $z$ direction, so that a particle that leaves the system through the right or left boundary (see Fig.~\ref{fig:sim_mod}) appears on the opposite boundary. Thus, it is not necessary to make the system very long in order to handle the whole particle trajectory from entering to ejection. The current sheet half-thickness, $L_x$, is equal to 1\oneunit{m}, while the width of the whole simulation region along $x$ is chosen to be 20\oneunit{m}, in order to avoid any influence of the boundaries on the particles inside the RCS. Plasma is continuously injected from the $x = \pm10\oneunit{m}$ sides of the simulation region with the rate $n E_{y0}/B_{z0}$.

In order to avoid numerical instabilities in the PIC method, the following constrains need to be satisfied:
\begin{eqnarray}
    c \Delta t < \Delta \xi, \\
    \Delta t < 0.2 \omega_{pe}^{-1}, \\
    \Delta \xi < \lambda_D,
\end{eqnarray}
where $\Delta t$ is the time step, $\Delta \xi$ is the grid step in any direction, $c$ is the speed of light, $\omega_{pe} = (4 \pi n e^2 / m_e)^{1/2}$ is the electron plasma frequency and $\lambda_D = (k T / 4 \pi n e^2)^{1/2}$ is the Debye length. To satisfy these conditions while keeping the code running time at a reasonable level, we use a reduced plasma number density $n=10^{10} \permcub$ in our simulations. Also the proton-to-electron mass ratio is reduced to $m_p/m_e=100$ in order to keep the proton acceleration time within reasonable computational limits. The spatial simulation grid has from $10$ to $100$ cells in the $z$ direction and $100$ cells in the $x$ direction with $\Delta z = \lambda_D$, $\Delta x = \lambda_D/5$ and $100$ superparticles per cell in average, while each superparticle represents $2\tento{7}$ real particles. The time step is $6\tento{-10}\oneunit{s}$.

As it was mentioned in Sec.~\ref{sec:intro}, we assume the existence of a background plasma with the real coronal density of $10^{16}\permcub$, and which is not a part of PIC simulation, due to the limits of a current power even in large computer clusters. This assumption is required to provide the background current $j_y = -1/\mu_0 \cdot dB_z/dx$ to sustain the equilibrium with the background magnetic configuration given by Eqs.~(\ref{eq:bz}-\ref{eq:by}), that will produce electron velocities $V_y = 1/(\mu_0ne) \cdot dB_z/dx$. For the real coronal plasma density the electron velocity which forms the equilibrium background current is $5\tento{5}\oneunit{m/s}$. This velocity is a few orders of magnitude lower than the velocities gained by the electrons at acceleration in the drifted electric field, thus it will not change significantly the electron distributions and is neglected in the current approach. By doing so we, possibly, eliminate only some low frequency instabilities (like ion-sound wave), which are not a part of our present study. However, this approach allows us to investigate the acceleration of a smaller number of particles ($n=10^{10} \permcub$) accelerated by the drifted electric field by taking into account the plasma feedback to these particles motion.

The simulations are supposed to run until a quasi-stationary state is reached. This running time corresponds to the acceleration time of the slowest particles (protons) and is usually less than $10^{-3} \oneunit{s}$. Also, since we are interested in the stationary state, and the background magnetic field in assumed to be in equilibrium with the the high density plasma, the initial state of the simulated plasma particles is not important. In practice, all particles that are initially present in the simulation region would be accelerated, ejected and replaced by the injected particles before the stationary state is reached. Thus, to speed up the simulation we do not initially have any particles in the simulation region. The injected particles are assumed to have a thermal distribution with the typical coronal temperature of $10^6\oneunit{K}$.

\subsection{Comparison with the test particle simulations}

The PIC simulations proved that the RCS magnetic configuration may accelerate plasma particles even if the plasma feedback is taken into account. The results of simulation obtained for an extremely low number density of particles ($10^{7} \permcub$) are plotted in Figs.~\ref{fig:vz_x_lownp} and \ref{fig:vz_x_lowne}. This low density simulation is only performed to verify that the PIC code can repeat the results obtained in the TP simulations (see Fig.~\ref{fig:vz_x}). Indeed, the formation of beams of accelerated protons (Fig.~\ref{fig:vz_x_lownp}) and electrons (Fig.~\ref{fig:vz_x_lowne}) is clearly seen. Similar to the TP simulations, the electrons and protons are ejected into the opposite semispaces with respect to the midplane ($x=0$).

The protons are unmagnetised inside the current sheet, so the ejection velocity of the "bounced" and "transit" protons are almost the same. On the other hand, the electrons are much more magnetised. Hence, similarly to what was shown in the TP simulations, the "bounced" electrons are not able to reach the midplane and, thus, they gain less energy than the "transit" electrons. The electric field $\tilde{E}_x$ caused by the charge separation is quite weak in this case, about 17\oneunit{V/m}.

In our further simulations the plasma density is accepted to be higher ($10^{10} \permcub$), but it is still several orders of magnitude lower than the typical coronal density ($10^{15} - 10^{16} \permcub$), which is unreachable due to the computational limitations. The higher density simulations (Figs.~\ref{fig:vz_x_hinp} and \ref{fig:vz_x_hine}) reveal more differences between the TP and PIC approaches. Proton dynamics in PIC (Fig.~\ref{fig:vz_x_hinp}) has not changed much because of the increased density, they are still ejected mainly into one semispace ($x<0$) with respect to the midplane. The acceleration rate of protons coincides with the theoretical value given by Eq.~(\ref{eq:enrg_x_unmag}), and their trajectories are close to those obtained from the TP simulations (see Fig.~\ref{fig:vz_x}). However, the proton orbit at the acceleration phase is wider ($\sim3\oneunit{m}$ versus $\sim1\oneunit{m}$ in Fig.~\ref{fig:vz_x}). Also, near the midplane the protons form the structure with a higher density on the sides and in the centre (Fig.~\ref{fig:vz_x_hinp}). A similar structure in the proton density inside the RCS was obtained by \citet{Zharkova08}.

\begin{figure}
  \centering
  \includegraphics[width=0.48\textwidth]{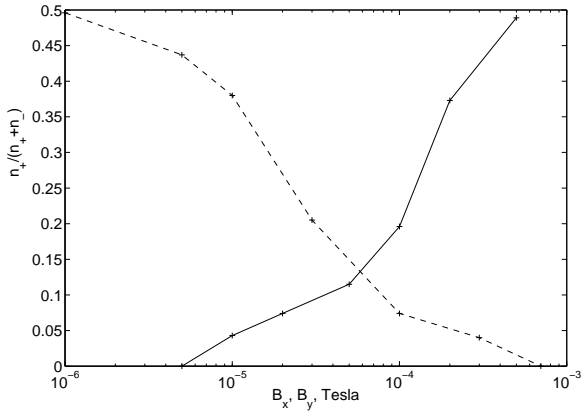}
  \caption{The fraction of protons that are ejected to the $x>0$ semispace as a function of $B_{y0}$ (dashed line, $B_{x0} = 2 \tento{-5} \oneunit{T}$) and $B_{x0}$ (solid line, $B_{y0} = 10^{-4} \oneunit{T}$). All other parameters of the current sheet are the same as in Fig.~\ref{fig:vz_x_hinp}.}
  \label{fig:separ}
\end{figure}

In addition there is a small number of protons that are ejected to the $x>0$ semispace (Fig.~\ref{fig:vz_x_hinp}), where electrons are normally ejected in TP simulations. Plot in Fig.~\ref{fig:separ} shows the fraction of the protons that are ejected to the $x>0$ semispace as a function of $B_{x0}$ and $B_{y0}$. The dependence on $B_{y0}$ qualitatively coincides with the asymmetry rate dependence obtained by \citet{Zharkova04}. In particular, for a very small $B_{y0}$ in both PIC and TP simulations the protons are ejected symmetrically with respect to the midplane. However, in PIC simulation all the protons are ejected to the one semispace ($x<0$) when $B_{y0} > 7 \tento{-4} \oneunit{T} \approx B_{z0}$, while \citet{Zharkova04} have shown that in the TP simulation accelerated particles are fully separated if $B_{y0}>1.5 \tento{-2} B_{z0}$. Thus, the particle trajectories have smaller asymmetry in PIC simulations compared to the TP ones.

The electron dynamics in PIC simulations, as it is seen from Fig.~\ref{fig:vz_x_hine}, is different from what was predicted by the TP simulations. Instead of a formation of the beams with a narrow energy distribution, the accelerated electrons are found to gain a wide range energy spectrum. Moreover, the electrons become ejected mainly into the same side as protons. Also the energies of ejected electrons in PIC simulations are smaller than in the TP simulations. In order to understand these new effects, let us study the electron and proton dynamics in more details by reconstructing the trajectories of particles in the RCS.

\begin{figure}
  \centering
  \includegraphics[width=0.48\textwidth]{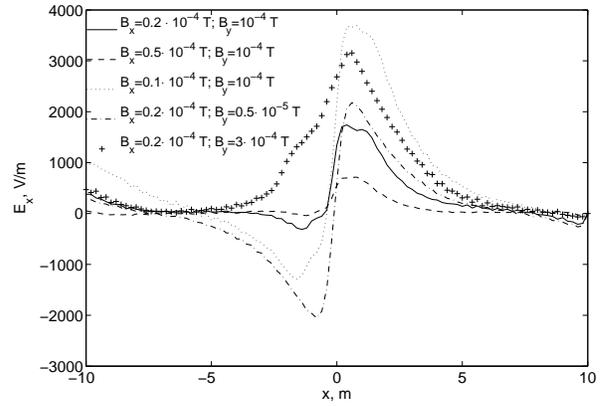}
  \caption{Electric field $\tilde{E}_x$ induced by particles in the PIC simulations for different values of $B_{x0}$ and $B_{y0}$. All other parameters of the current sheet are the same as in Fig.~\ref{fig:vz_x_hinp}.}
  \label{fig:E_x}
\end{figure}

\begin{figure}
  \centering
  \includegraphics[width=0.48\textwidth]{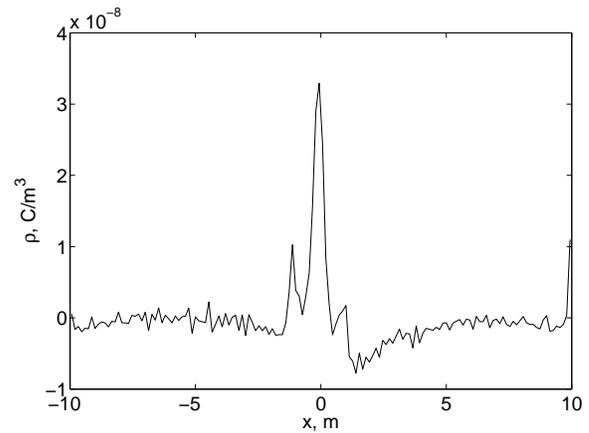}
  \caption{Charge density $\rho$ as a function of $x$ deduced from the PIC simulation. Current sheet parameters are the same as in Fig.~\ref{fig:vz_x_hinp}.}
  \label{fig:Rho_x}
\end{figure}

\begin{figure*}
  \centering
  \subfloat[Two protons, injected at $x=\pm10 \oneunit{m}$ with zero initial velocities.]{
    \includegraphics[width=0.48\textwidth]{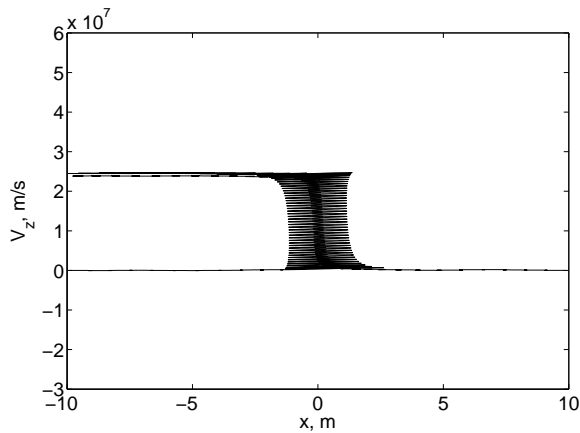}
    \label{fig:tr_p}}
  \subfloat[Electron, injected at $x=-10 \oneunit{m}$ with initial velocity $V_z=-5 \tento{6} \oneunit{m/s}$.]{
    \includegraphics[width=0.48\textwidth]{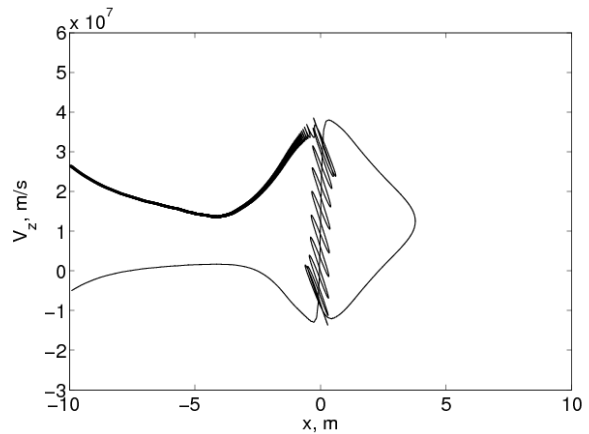}
    \label{fig:tr_e1}}\\
  \subfloat[Electron, injected at $x=10 \oneunit{m}$ with initial velocity $V_z=-1 \tento{7} \oneunit{m/s}$.]{
    \includegraphics[width=0.48\textwidth]{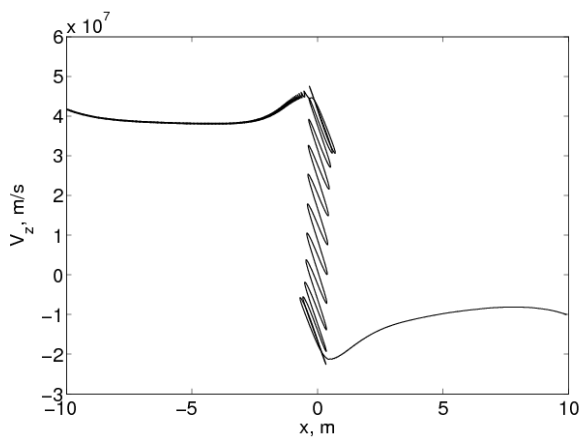}
    \label{fig:tr_e2}}
  \subfloat[Electron, injected at $x=-10 \oneunit{m}$ with initial velocity $V_z=-5 \tento{6} \oneunit{m/s}$.]{
    \includegraphics[width=0.48\textwidth]{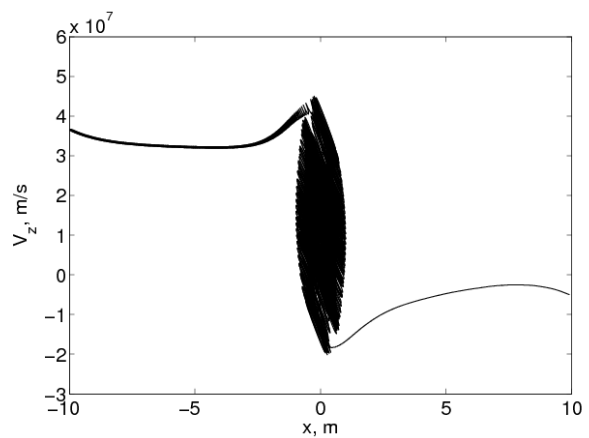}
    \label{fig:tr_e3}}
  \caption{Trajectories of particles on the $V_z$-$x$ phase plane during acceleration inside a current sheet. The trajectories were obtained in the TP simulations, with the electric field $\tilde{E}_x(x)$ taken from the PIC simulation (Fig.~\ref{fig:E_x} solid line). Current sheet parameters are the same as in Fig.~\ref{fig:vz_x_hinp}.}
\end{figure*}

\subsection{Induced electric field and particle trajectories}

The differences between the TP and PIC simulations could only occur because of the additional electric and magnetic fields induced locally by the accelerated particles. The PIC simulations has shown that the induced magnetic field $\mathbf{\tilde{B}}$ is much smaller than the background $\mathbf{B}$. On the other hand, the induced electric field $\mathbf{\tilde{E}}$ is essential. Its absolute value is even larger than the background (drifted) field $E_y$ induced by a magnetic reconnection (see Eq.~(\ref{eq:ey})).

Since our 2D system is invariant in the $y$ dimension there is no charge separation in this direction. Thus the induced $\tilde{E}_y$ can be caused only by the time variation of the magnetic field and is found to be small. In this subsection we consider the electric field $\tilde{E}_x$ which is perpendicular to the current sheet. Fig.~\ref{fig:E_x} is a plot of $\tilde{E}_x$ as a function of $x$ averaged over the $z$ coordinate for different values of $B_{x0}$ and $B_{y0}$. This field appears due to separation of the electrons and protons across the current sheet, which leads to a local non-neutrality of the plasma. The field becomes stronger when $B_{x0}$ decreases or $B_{y0}$ increases. Also, when $B_{y0} \rightarrow 0$ the electric field $\tilde{E}_x$ and the system itself become symmetric. The distribution of a charge density $\rho(x)$, which generates the polarisation field $\tilde{E}_x(x)$, is shown in Fig.~\ref{fig:Rho_x}. Note, that the average charge density over the simulation region is close to zero, which mean that the current sheet as a whole remains electrically neutral.

In order to reconstruct the particle trajectories, we use the TP code, where the induced electric field $\tilde{E}_x$ obtained from PIC is added to the background electro-magnetic configuration given by Eqs.~(\ref{eq:bz}-\ref{eq:ey}). The trajectories of the two protons in the $x$-$V_z$ phase plane is shown in Fig.~\ref{fig:tr_p}. These trajectories are very similar to those obtained in the TP simulations without $\tilde{E}_x$ (see Fig.~\ref{fig:vz_x}). The only difference is that during the acceleration phase the "bounced" proton has a wider orbit. It is clear now that this wide orbit is responsible for the two smaller peaks at about $\pm 1 \oneunit{m}$ in the charge density plot (Fig.~\ref{fig:Rho_x}). On the other hand, the narrow orbit of the "transit" proton forms the central peak in this plot.

\begin{figure*}
  \centering
  \subfloat[$B_{x0}=2 \tento{-5} \oneunit{T}$, $B_{y0}=5 \tento{-6} \oneunit{T}$]{
    \includegraphics[width=0.35\textwidth]{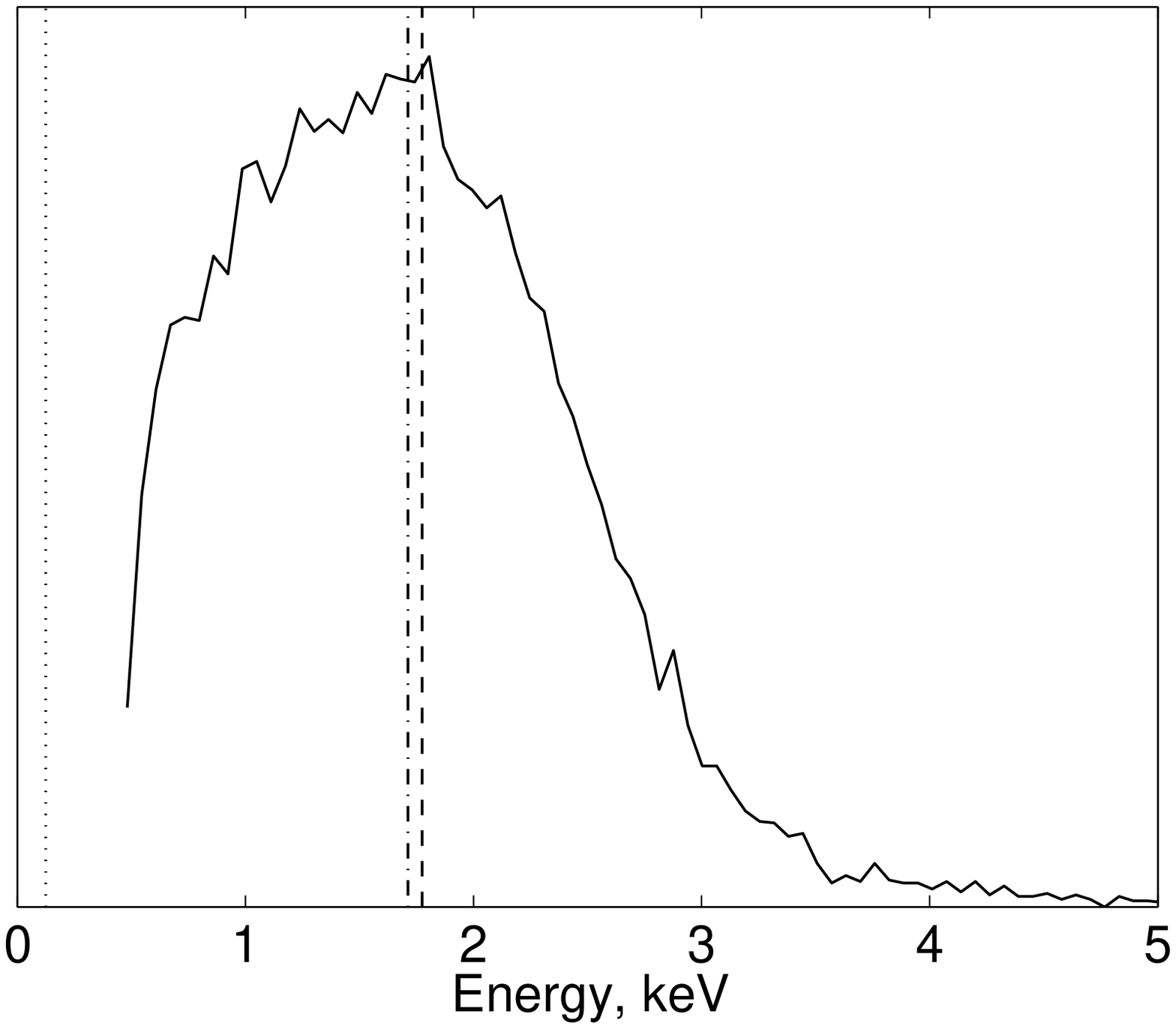}
    \label{fig:en24}}\qquad\qquad\qquad\qquad
  \subfloat[$B_{x0}=2 \tento{-5} \oneunit{T}$, $B_{y0}=3 \tento{-4} \oneunit{T}$]{
    \includegraphics[width=0.35\textwidth]{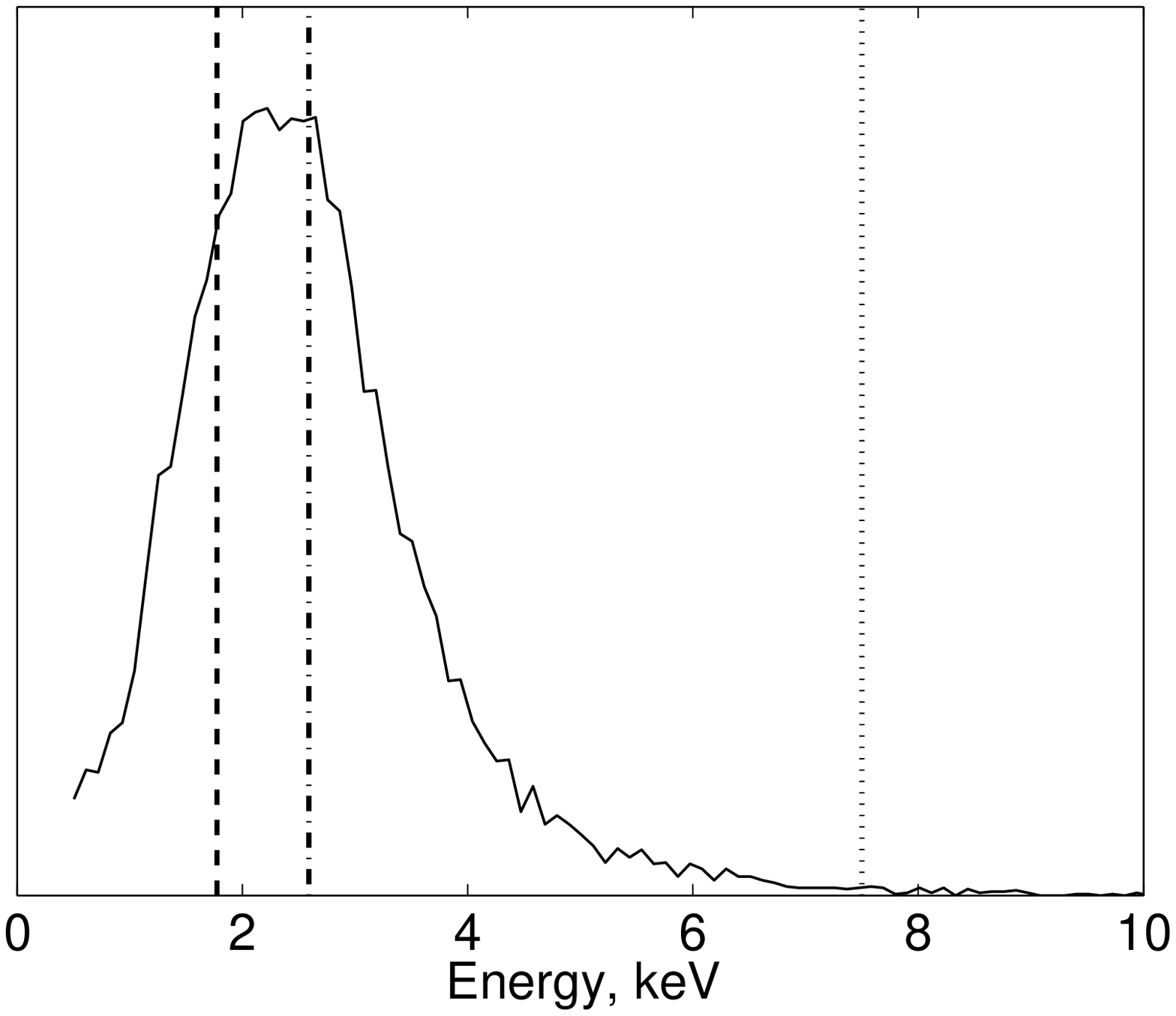}
    \label{fig:en18}}\\
  \subfloat[$B_{x0}=10^{-4} \oneunit{T}$, $B_{y0}=10^{-4} \oneunit{T}$]{
    \includegraphics[width=0.35\textwidth]{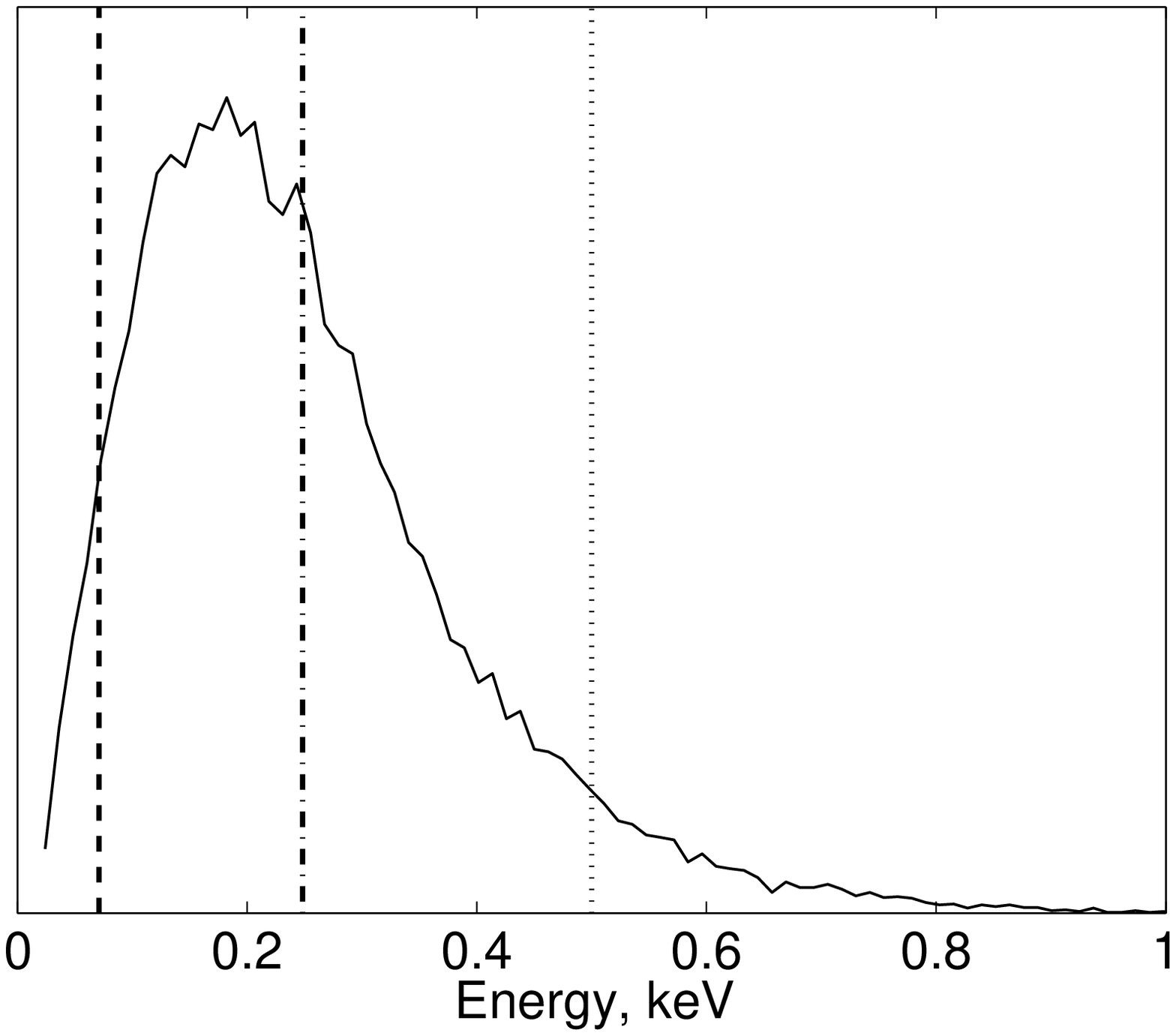}
    \label{fig:en07}}\qquad\qquad\qquad\qquad
  \subfloat[$B_{x0}=5 \tento{-5} \oneunit{T}$, $B_{y0}=10^{-4} \oneunit{T}$]{
    \includegraphics[width=0.35\textwidth]{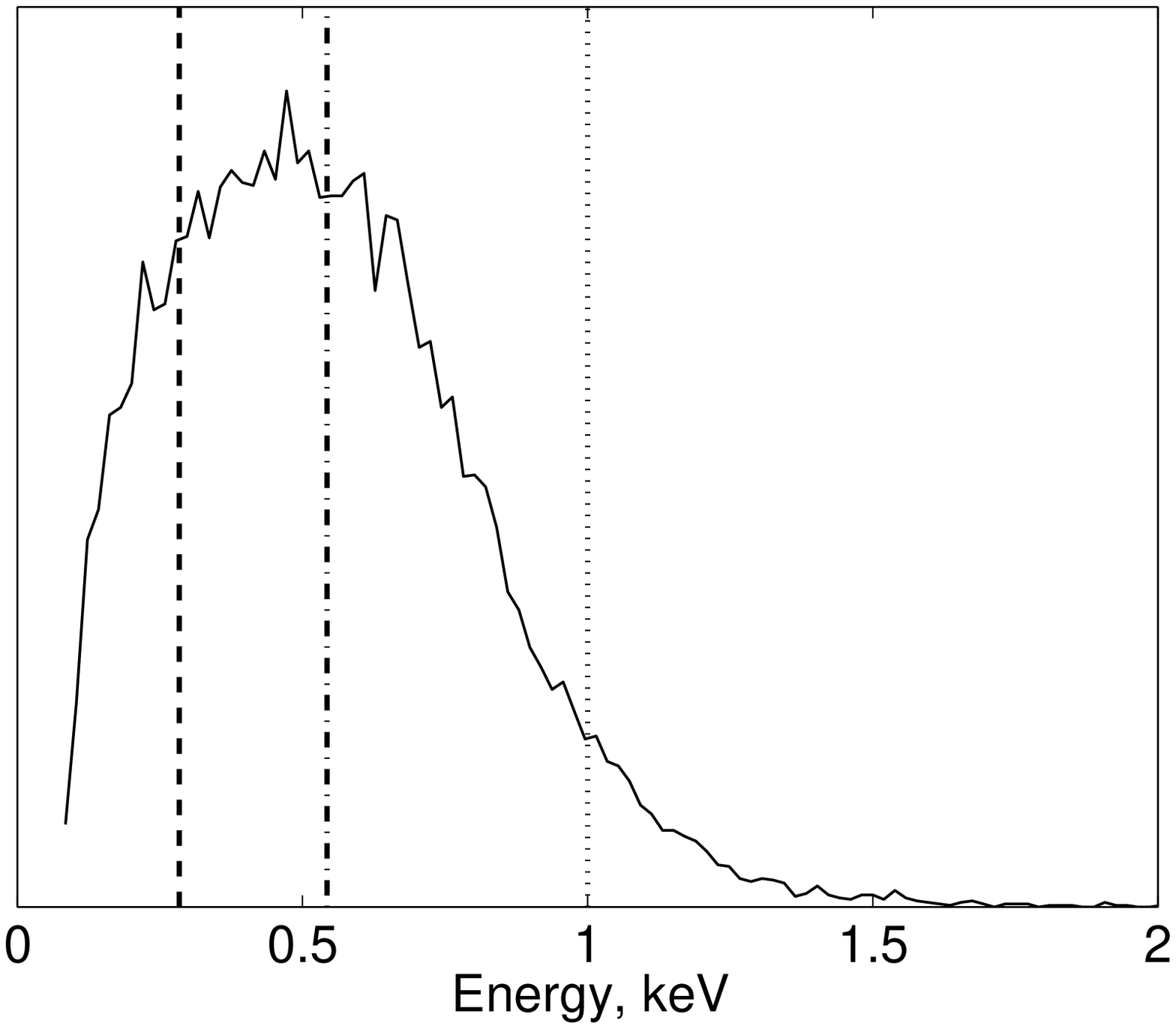}
    \label{fig:en22}}
  \caption{The energy spectra of the ejected electrons. Dashed line -- Eq.~(\ref{eq:enrg_x_unmag}), dotted line -- Eq.~(\ref{eq:enrg_x_mag}), dash-dotted line -- mean energy of the ejected electrons. Current sheet parameters are the same as in Fig.~\ref{fig:vz_x_hinp}.}
  \label{fig:energy}
\end{figure*}

The trajectories of electrons are much more complicated. Firstly, let us consider an electron that enters from the $x<0$ semispace (Fig.~\ref{fig:tr_e1}). The dynamics of this electron is similar to the dynamics of the "transit" electron in Fig.~\ref{fig:vz_x} -- the electron drifts towards the midplane, becomes accelerated and ejected to the $x>0$ semispace. However, the polarisation field $\tilde{E}_x(x)$, which extends beyond the current sheet and has a component parallel to the magnetic field, decelerates the ejected electron. For the chosen magnitudes of $B_x$ and $B_y$ the majority of electrons can not escape to the $x>0$ semispace, instead they are dragged back towards the current sheet and become indistinguishable from the electrons entered from the $x>0$ semispace.

The electrons that come from the positive $x$ side demonstrate a rather different dynamics in comparison with the case when $\tilde{E}_x = 0$. It turns out that the electron, which is "bounced" from RCS in the absence of $\tilde{E}_x$, can now reach its midplane. In the vicinity of the midplane, the electron becomes unmagnetised and oscillates with the gyrofrequency determined by $B_y$ (Fig.~\ref{fig:tr_e2}) and, after some time of oscillation, the electron is ejected. If the electron initial velocity is small, it can be quasi-trapped inside the RCS (Fig.~\ref{fig:tr_e3}). Such electron is accelerated on the midplane, ejected from it, then decelerated outside the RCS and returns back to the midplane. This cycle is repeated for several times, until it finally gains enough energy to escape the RCS. Since the magnitude of the polarisation field $\tilde{E}_x(x)$ is smaller at $x<0$ than at $x>0$ (see Fig.~\ref{fig:E_x}), it is easier for the electron to escape to the $x<0$ semispace. Thus, most of the electrons are ejected to the same semispace as protons.

\subsection{Energy spectra}

The energy distributions of accelerated electrons for different magnitudes of transverse, $B_{x0}$, and guiding, $B_{y0}$, magnetic components are calculated for those electrons outside the RCS in the negative semispace ($-10\oneunit{m} < x < -5 \oneunit{m}$) and plotted in Fig.~\ref{fig:energy}. Only escaping electrons are taken into account, i.e. the electrons for which the velocity component along the magnetic field projected on the $x$ axis is larger than the $\mathbf{E} \times \mathbf{B}$ drift velocity projected on the $x$ axis. For the accepted configuration and in the limit of a small $B_x$, this corresponds to the electrons with $V_z > E_{y0}/B_{x0}$.

It can be seen from Fig.~\ref{fig:energy} that for any $B_{x0}$ and $B_{y0}$ the ejected electrons form a wide single-peak energy distribution with a width of the order of the mean energy. This is contrary to the TP simulations and the lower density PIC simulations, where the two narrow energy electron beams are formed (see Fig.~\ref{fig:vz_x} for the TP simulation and Fig.~\ref{fig:vz_x_lowne} for the lower density PIC simulation). The presence of the polarisation electric field shifts the high energy peak towards the low energy one, while the latter is expanded.

If the guiding field, $B_y$, is negligibly small (Fig.~\ref{fig:en24}), then the mean energy of the ejected electrons coincides with the analytical value given by Eq.~(\ref{eq:enrg_x_unmag}). For the stronger guiding field (Fig.~\ref{fig:en18}) the mean energy is somewhat higher than the energy of the slow beam in the TP simulation given by Eq.~(\ref{eq:enrg_x_unmag}), and substantially lower than the energy of the fast beam given by Eq.~(\ref{eq:enrg_x_mag}).

The energy spectra dependence on $B_x$ is shown in Figs.~\ref{fig:en07} and \ref{fig:en22}. They indicate that the mean energy strongly depends on $B_x$ as it is predicted by Eqs.~(\ref{eq:enrg_x_unmag}) and (\ref{eq:enrg_x_mag}). The lower $B_x$ the higher their mean energy and the wider the energy distribution.

\subsection{Generation of the Langmuir waves}

\begin{figure}
  \centering
  \includegraphics[width=0.5\textwidth]{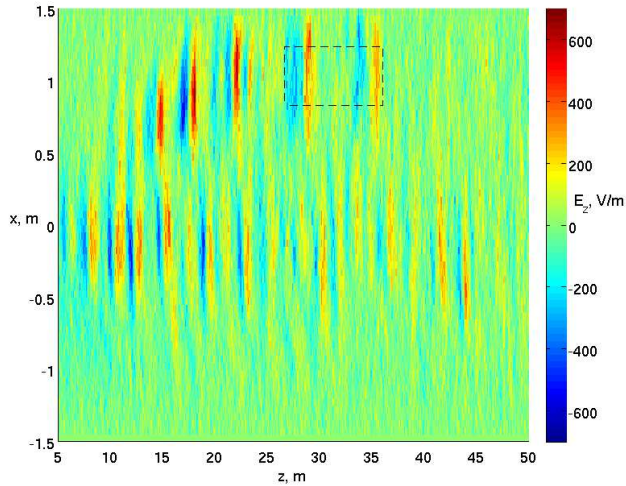}
  \caption{Electric field $\tilde{E}_z$ induced by particles in PIC simulation ($B_{z0} = 10^{-3} \oneunit{T}$, $B_{y0} = 10^{-4} \oneunit{T}$, $B_{x0} = 4 \tento{-5} \oneunit{T}$, $E_{y0} = 250 \oneunit{V/m}$, $m_p/m_e=10$, $n = 10^{12} \permcub$).}
  \label{fig:E_z}
\end{figure}

\begin{figure}
  \centering
  \includegraphics[width=0.45\textwidth]{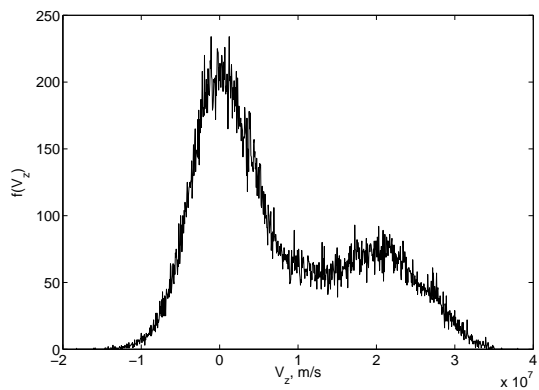}
  \caption{Distribution function of electrons calculated for the region indicated by the dashed rectangle in Fig.~\ref{fig:E_z}.}
  \label{fig:distr}
\end{figure}

Another interesting effect observed in the PIC simulations is the excitation of waves induced by accelerated particles. As it can be seen from Fig.~\ref{fig:E_z}, the $\tilde{E}_z$ component of the induced electric field is structured with a characteristic length scale of about $\lambda_{wave} \approx 2 \oneunit{m}$. This structure propagates in time in the positive direction of the $z$ axis. The speed of this propagation is about $V_{wave} \approx 1.3 \tento{7} \oneunit{m/s}$, which makes the temporal period of oscillations about $T_{wave} \approx 1.5 \tento{-7} \oneunit{s}$. Since the plasma frequency for $n = 10^{12} \permcub$ is $9.1 \tento{6} \oneunit{Hz}$, the generated wave is the Langmuir wave. Note also that the oscillating component of the excited wave $\tilde{E}_z$ is parallel to the direction of propagation, which corresponds to the polarisation of the Langmuir wave.

In order to clarify the mechanism of instability, let us plot a $V_z$ distribution function of electrons in the region where the instability occurs (this region is indicated by the dashed rectangle in Fig.~\ref{fig:E_z}). Fig.~\ref{fig:distr} indicates that the electrons have the typical "bump-in-tail" unstable distribution. The range of velocities $V_z$ for which the derivative $df/dV_z$ is positive is from $1.3 \tento{7}$ to $2 \tento{7} \oneunit{m/s}$, these values correspond to the phase velocity of the Langmuir wave $V_{wave}$ deduced earlier.

The generation of the Langmuir waves is not the only possible instability in an RCS. For example, \citet{Drake94} have shown the possibility of whistles waves excitation in thin current layers. However, in our simulations we can observe only waves with length scales less than the size of the simulation domain. Also, since the current simulation system is 2D in space (and 3D in velocities), the waves which are essentially 3D cannot be generated. The investigation of other instabilities, apart from Langmuir waves, will be a subject of the forthcoming study.

\section{Conclusions and discussions}

In this paper we present the PIC simulations combined with the TP approach for particle acceleration in a reconnecting current sheet with given magnetic field topology. The acceleration is studied in a small part of the current sheet with a 3D background magnetic field and a drifted electric field induced by the magnetic diffusion. The simulations are carried out for thin RCS of $2 \oneunit{m}$. The main (reconnecting) component of the magnetic field is $10^{-3} \oneunit{T}$, while the other two components are varied in wide range.

The background magnetic and electric fields are considered to be stationary while varying across the current sheet and being constant in other dimensions. The PIC simulations is used to study particle acceleration in the combined background electro-magnetic fields and those fields induced by the accelerated particles. The TP approach is used to study how these additional fields affect the particle trajectories inside the 3D RCS, that, in turn, helps to understand the PIC results.

The PIC simulations carried out for a lower plasma density ($10^7 \permcub$) reveal the same particle trajectories as in the TP approach. In particular, for the RCS with lower density we confirm the separation, or asymmetry, of particle beams with the opposite charges, which are ejected into the opposite semispaces with respect to an RCS midplane. For a higher plasma density ($10^{10} \permcub$), the accelerated particles are found to produce the additional electric fields: a polarisation field, $\tilde{E}_x$, caused by the charge separation of accelerated particles across the current sheet and by a turbulent field of Langmuir waves, $\tilde{E}_z$, caused by the "bump-in-tail" energy spectra of the accelerated electrons.

The magnitude of the polarisation field $\tilde{E}_x$ is shown to be very high that leads to essential modification of the trajectories of electrons, and, to some extent, those of protons. There is a difference in the acceleration of the "transit" particles (those which enter from and are ejected to the opposite semispaces towards the midplane) and "bounced" particles (those which enter from and are ejected to the same semispace towards the midplane).

In particular, during their acceleration, the "bounced" protons are found to have wider orbits than the "transit" ones. It was also found in PIC simulations that the protons are ejected less asymmetrically with respect to the midplane compared to the trajectories found in the TP simulations. For example, in the TP approach the proton ejection is fully asymmetric (all protons are ejected to the same semispace) if the guiding field $B_{y0}$ is larger than $1.5 \tento{-2} B_{z0}$, while in the PIC simulations this critical guiding field is of the same order of magnitude of the main component $B_{z0}$.

In the current RCS topology, if there is an asymmetry of the accelerated protons at ejection, the accelerated electrons in the PIC simulations are shown to be ejected to the same semispace as the protons, i.e. there is no non-local charge separation as predicted by the TP simulations \citep{Zharkova04}. The electrons, which are "bounced" in the TP approach, are in the PIC simulations dragged by the polarisation field $\tilde{E}_x$ to the midplane. At the same time, the "transit" electrons, which are ejected to the opposite semispace from protons in the TP approach, cannot leave the simulation region in PIC, i.e. they become dragged by the polarisation field $\tilde{E}_x$ back to the RCS. Since in PIC the electrons are eventually ejected into the same semispace as protons, the charge separation can only happen locally, about $10 \oneunit{m}$ around the midplane. This is valid for the small simulation region considered in the current paper, while the PIC simulations for the larger region could result in separation of high energy particles across the RCS.

The energy distribution of ejected electrons in the PIC simulations is found to be rather wide, i.e. the width of the distribution is of the order of magnitude of the mean energy, which is different from the energy spectra found in TP approach \citep{Zharkova04}. However, the mean energy of electrons is found to be consistent with the TP simulation and the analytical estimations for a weak guiding field. In spite of the strong guiding field can significantly enhance the ejection energy of the "transit" electrons in the TP simulations, in the PIC simulations this enhancement is shown to be much less pronounced if the polarisation field  $\tilde{E}_x$ is taken into account.

PIC simulations reveal the additional periodic electric field $\tilde{E}_z$. This field is found to be formed by the waves generated by the beam of accelerated electrons with the unstable "bump-in-tail" velocity distribution. For the magnitudes of the background electric and magnetic fields considered in the current study, the phase speed of generated waves is $\approx 1.3 \tento{7} \oneunit{m/s}$, the wave length is $\approx 2 \oneunit{m}$ and the period of oscillations is $\approx 1.5 \tento{-7} \oneunit{s}$, which corresponds to the Langmuir wave frequency. The turbulent electric field $\tilde{E}_z$ of this waves is likely to be the reason of the wide energy distribution of the ejected electrons derived from the PIC simulation.

Therefore, we show that the feedback of the ambient plasma in an RCS in a form of electric fields produced by the accelerated particles is rather strong and modifies substantially the particle trajectories and energy spectra gained during their acceleration. These effects are essential even for a small fraction of the particles simulated by PIC in the present study. The full particle simulation for a real plasma density can reveal further effects in the plasma dynamics during a magnetic reconnection, that will be a scope of the forthcoming studies with the increased computer power.

\begin{acknowledgments}
This research is funded by the Science Technology and Facility Council (STFC) project PP/E001246/1. The computational work was carried out on the joint STFC and SFC (SRIF) funded cluster at the University of St Andrews, UK.
\end{acknowledgments}

\bibliography{bib_gen}

\end{document}